\documentclass[runningheads]{llncs}
\usepackage{graphicx}
\usepackage{amsmath}
\usepackage[colorlinks=true,
            linkcolor=blue,
            urlcolor=cyan,
            citecolor=green]{hyperref}
\usepackage{booktabs}
\usepackage{cite}
\usepackage{tabularx}

\title{New Kid in the Classroom: Exploring Student Perceptions of AI Coding Assistants}
 
\titlerunning{Exploring Student Perceptions of AI Coding Assistants}

\author{Sergio Rojas-Galeano\inst{1}}
\authorrunning{Sergio Rojas-Galeano}

\institute{Universidad Distrital Francisco José de Caldas, Bogotá, Colombia\\
\email{srojas@udistrital.edu.co}}

\begin{document}

\maketitle

\begin{abstract}
The arrival of AI coding assistants in educational settings presents a paradigm shift, introducing a ``new kid in the classroom'' for both students and instructors. Thus, understanding the perceptions of these key actors about this new dynamic is critical. This exploratory study contributes to this area by investigating how these tools are shaping the experiences of novice programmers in an introductory programming course. Through a two-part exam, we investigated student perceptions by first providing access to AI support for a programming task and then requiring an extension of the solution without it. We collected Likert-scale and open-ended responses from 20 students to understand their perceptions on the challenges they faced. Our findings reveal that students perceived AI tools as helpful for grasping code concepts and boosting their confidence during the initial development phase. However, a noticeable difficulty emerged when students were asked to work unaided, pointing to potential overreliance and gaps in foundational knowledge transfer. These insights highlight a critical need for new pedagogical approaches that integrate AI effectively while effectively enhancing core programming skills, rather than impersonating them.

\keywords{Artificial intelligence \and Programming education \and Code assistants \and Student performance \and Educational technology}
\end{abstract}

\section{Introduction}

AI coding assistants are fast becoming the ``new kid in the classroom'', fundamentally altering the landscape of education and creating a paradigm shift for both students and instructors. When deployed as tools like ChatGPT or Grok, these models can generate, debug, and explain code with increasing fluency, fundamentally changing how novice programmers learn and engage with their coursework. The integration of these powerful tools into an introductory programming curriculum raises a central question: What is the actual student experience when learning to program with AI support?

While these AI assistants offer undeniable benefits, such as instant feedback and assistance with common syntactic issues, their presence also introduces concerns about overreliance and reduced engagement with fundamental concepts. Most existing research has focused on professional development contexts or examined AI tools in isolation, leaving a critical gap in our understanding of how students themselves perceive and engage with these tools in an educational setting.

This exploratory study addresses that gap by taking a preliminary step into this new pedagogical territory. We investigate how students engage with these tools during a two-part programming assignment and how their perceptions shift when AI support is removed. By examining student attitudes and experiences, we aim to offer preliminary insights into the benefits and limitations of AI tools. The findings from this assessment are informative for designing pedagogical strategies that effectively integrate AI to promote genuine learning, moving beyond its use as a crutch for simply completing tasks.

\subsection*{Related Work}

The rapid integration of AI-powered code assistants into computer science education has prompted a wave of research exploring their impact on novice programmers. Several studies have focused on student interactions with these tools for specific tasks. For example, Alves and Cipriano \cite{alves2024give} analyzed student interaction with GPT models during object-oriented programming tasks, while Tian et al. \cite{tian2024teaching} similarly examined creative coding and conversational AI interactions with ChatGPT across experience levels, including beginners. Becker et al. \cite{becker2023generative} compared the performance of tools like ChatGPT to that of high-achieving students in an introductory course, raising implications for automated support. Similarly, G\"uner and Er \cite{guner2025ai} conducted a controlled experiment revealing distinct learner profiles and finding that students benefited most when using AI for code refinement rather than for full code generation.

Other research has explored pedagogical and design-focused applications. Fowler \cite{fowler2024strategies} has provided a broad synthesis of assessment strategies in early programming education, while Shynkar \cite{shynkar2023influence} has explored the effects of AI on learning trajectories. From a design perspective, Kulangara \cite{kulangara2024designing} and Luo et al. \cite{luo2025assessing} presented platforms for AI-supported education and personalized mentoring.

While this growing body of literature confirms that LLMs are reshaping the pedagogical landscape, much of it is situated in controlled lab settings or focuses on a single aspect of AI integration. Our study complements this existing body of work by addressing a critical gap: understanding the student experience in an authentic educational context. Instead of a structured lab environment, we explore how novices perceive and reflect on AI assistance during a real, two-part evaluative task. By examining how students engage with AI and later transfer their knowledge to an unaided task, our research moves the discussion beyond experimental findings to actual classroom practices, providing a deeper understanding of the challenges and opportunities of AI integration.

\section{Methodology}

\subsection{Research Context}
This study, conducted as a case study, involved 20 undergraduate students enrolled in an introductory Object-Oriented Programming (OOP) course at the District University of Bogota during the first semester of 2025. We aimed to understand the emerging dynamics of AI integration within a naturalistic educational setting, rather than a highly controlled experimental environment. All participants had prior programming experience, but their familiarity with AI tools varied.

\subsection{Study Protocol}
Our study followed a three-session protocol, conducted over three consecutive weeks in the Department's computing facilities, to observe student behavior and gather reflections on AI help during a two-part programming task.

\paragraph{Session 1: Programming with AI support.}
Students were tasked with developing a complete Java application from a medium-difficulty specification. To observe their initial interactions with AI, they were explicitly allowed to use any AI-powered code assistant of their choice (e.g., ChatGPT, Perplexity, Grok, DeepSeek), along with the BlueJ IDE and official Java API documentation. The 90-minute session was conducted individually under supervision.

\paragraph{Session 2: Programming without AI support.}
One week later, students returned for a second session focused on a knowledge transfer task. Each student was assigned a short extension to their original application, designed to require simple extensions to their existing code. During this 30-minute session, students were prohibited from using any AI assistance. They worked individually with access only to the BlueJ IDE and Java API documentation, allowing us to observe how they behave without AI support.

\paragraph{Session 3: Post-Task Reflections.}
In the final session, students completed an anonymous survey to capture their experiences and perceptions on AI use. The survey included both Likert-scale items and open-ended questions designed to capture rich, qualitative data about the challenges they faced.

\subsection{Study Design and Data Collection}

This study employed a sequential, two-phase protocol in which all participants experienced both conditions of AI assistance (With AI / Without AI). The design was chosen to provide comparative insights into how student perceptions and behaviors shifted as they moved from an AI-supported task to an independent, unaided task. The primary objective was not to establish a causal relationship or measure program correctness; instead, the emphasis was on gaining a deeper understanding of the student experience.

For this purpose, we collected qualitative observations of student interactions with AI tools during the first programming task, as well as of their knowledge transfer during the second unaided task. Additionally, we gathered self-reported perceptions and attitudes toward AI through an anonymous post-study survey, completed after the second programming session. Since no personal or sensitive data were collected and participants provided informed consent, no formal ethical approval was required. The survey included (for further details, see Appendix~\ref{appendix:survey}):

\begin{itemize}
\item[\textbullet] \textbf{Multiple choice and Likert-scale items} assessing students’ programming proficiency, AI tool usage, and perceived effects of AI on task-solving (e.g., speedup, understanding, and solution variety).
\item[\textbullet] \textbf{Open-ended questions} exploring students’ experiences in greater depth, including their understanding and adaptation of AI-generated code, perceived difficulties, general views of the exam, and how confident they felt when using AI assistance.
\end{itemize}

\subsection{Data Analysis}

Data analysis followed a mixed-methods approach to explore student perceptions and experiences. We used descriptive statistics to summarize AI usage and student perceptions from the Likert-scale items. To complement this, we explored potential associations among key variables.

Open-ended responses, which form the core of this study, were subjected to a thematic analysis to identify recurring patterns and extract meaningful insights into students’ experiences with and without AI support. All statistical analyses and visualisations were performed using the latest versions of \texttt{Python} and its scientific libraries, including \texttt{pandas}, \texttt{numpy}, \texttt{scipy}, and \texttt{matplotlib}.

\section{Results}

\subsection{AI Usage Characterisation}

\begin{figure}[t]
\centering
\includegraphics[width=.9\textwidth]{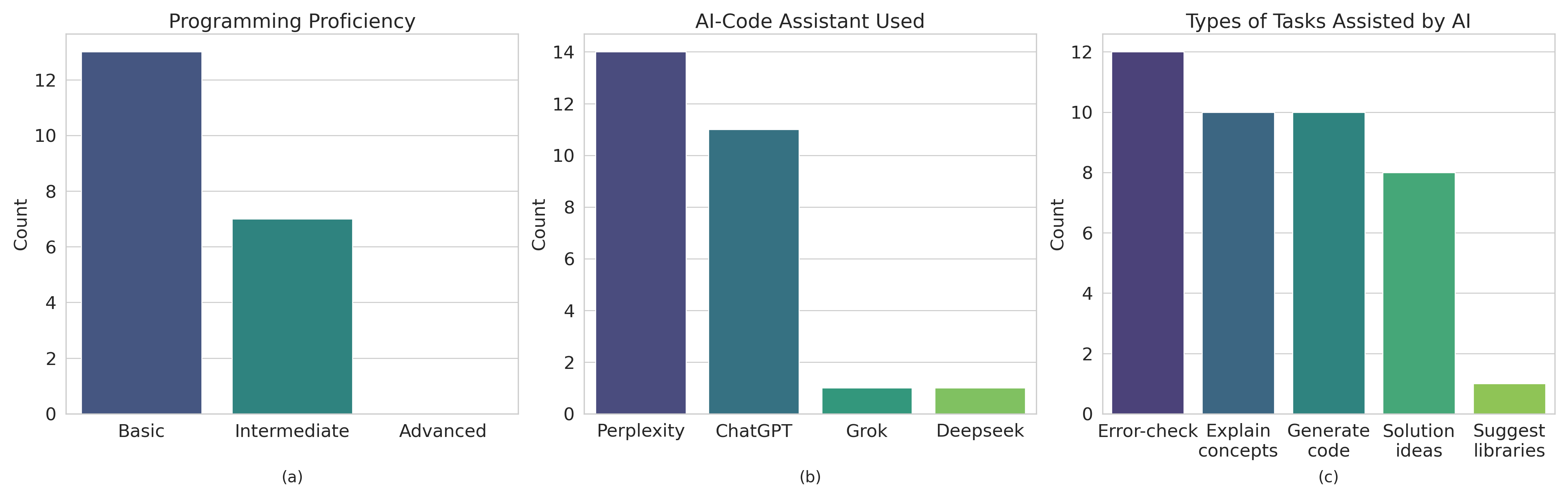}
\caption{Summary of participant profile and AI usage characterisation: (a) self-reported programming level, (b) preferred AI code assistants, and (c) common AI use cases. Multiple responses were allowed for (b) and (c).}
\label{fig:characterisation}
\end{figure}

Figure~\ref{fig:characterisation} summarises the cohort's profile and AI-related practices during the exam (based on survey questions~Q1-Q3,  Appendix~\ref{appendix:survey}). As panels (a), (b) and (c) show, the majority of students self-identified as basic-level programmers, with a smaller proportion indicating intermediate proficiency; none considered themselves advanced. Perplexity appears as the most commonly used AI assistant, followed by ChatGPT, while mentions of Grok and Deepseek were isolated. When asked about their main purposes for using AI tools, students most frequently cited debugging tasks, along with code explanation and generation. This characterisation hints at the cohort's novice background and highlights a strong reliance on AI for support, particularly for coding and debugging, which provides essential context for interpreting their performance across the two exam sessions.

\begin{figure}[h!]
    \centering
    \includegraphics[width=.8\textwidth]{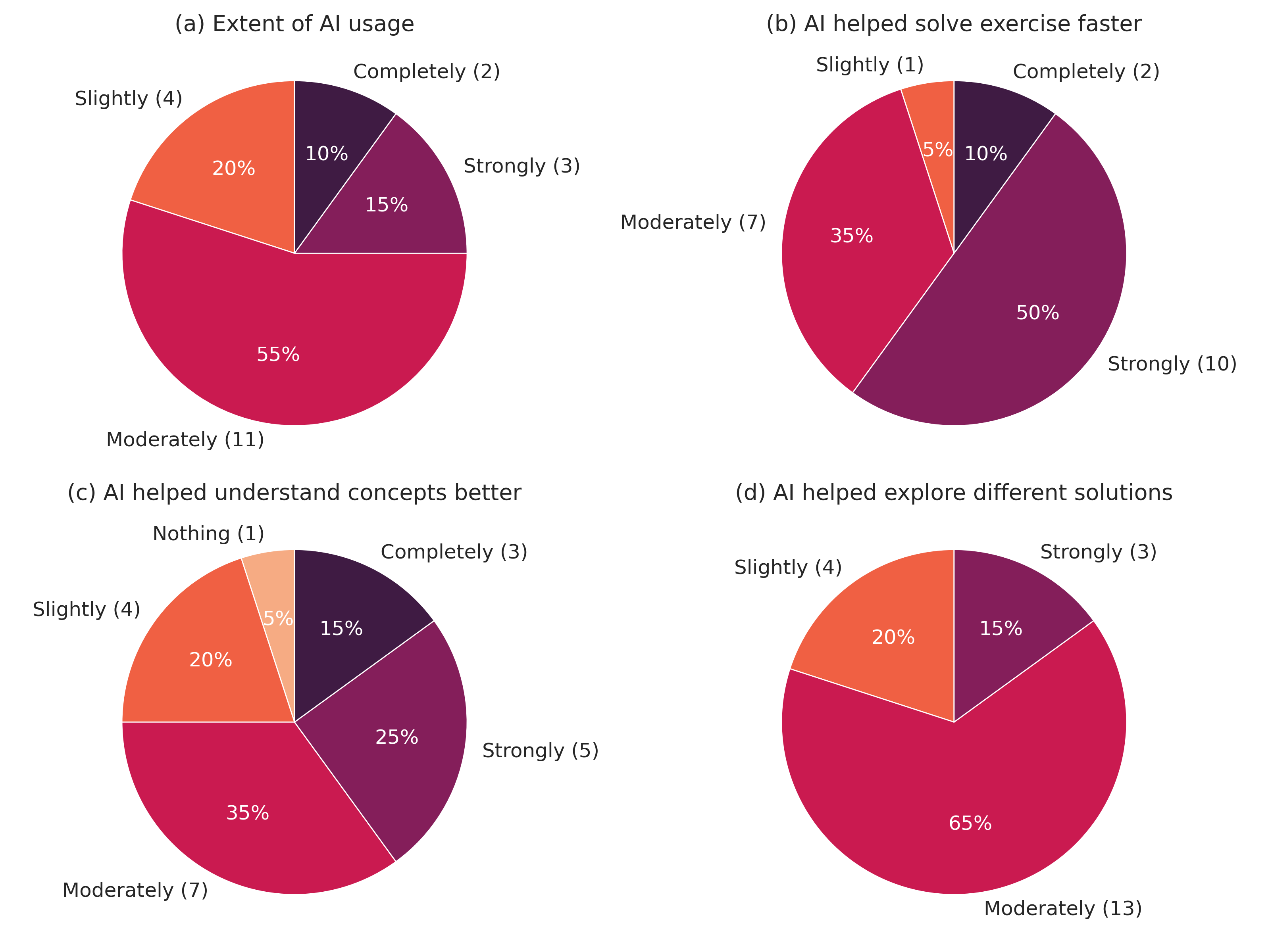}
    \caption{Student perceptions of AI usage and benefits during the first part of the exam. Each panel shows the distribution of responses to one of four Likert-scale prompts (Q4–Q7), rated from ``Nothing'' to ``Completely'': (a) extent of AI usage, (b) perceived gains in problem-solving speed, (c) improved conceptual understanding, and (d) enhanced exploration of alternative solutions.}
    \label{fig:likert-pies}
\end{figure}

\subsection{Perceived Benefits of AI Assistance}

Students responded to four questions (Q4–Q7, Appendix~\ref{appendix:survey}) regarding their experience with AI during the first part of the exam, using a five-point Likert scale from \textit{Nothing} to \textit{Completely} While the pie charts in Figure~\ref{fig:likert-pies} summarise the raw distribution of these categorical responses, a numerical mapping (1–5) was applied solely for descriptive analysis.

Overall AI usage averaged $\mu = 3.15$ ($\sigma = 0.88$), suggesting moderate engagement. The highest ratings were given to perceived gains in speed ($\mu = 3.65$, median = 4), indicating that most students believed AI helped them complete tasks more efficiently.

Ratings for improved conceptual understanding were similarly moderate ($\mu = 3.25$), though with greater variability ($\sigma = 1.12$), suggesting divergent experiences. Perceptions of AI aiding solution exploration were more neutral ($\mu = 2.95$), with students less consistently viewing it as a source of creative support.

\begin{figure}[t]
    \centering
    \includegraphics[width=0.5\textwidth]{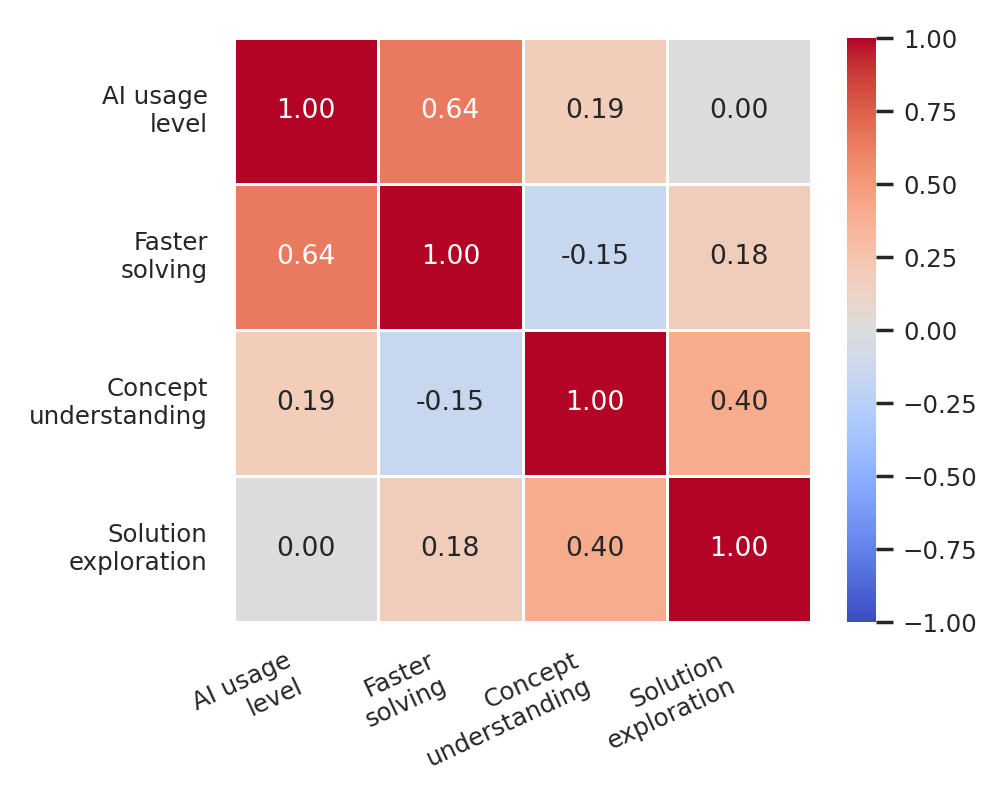}
    \caption{Spearman correlations between perceived AI usage (Q4) and reported benefits (Q5–Q7). Only the link between usage and perceived speed gains was statistically significant ($p < 0.01$).}
    \label{fig:correlation-q4-q7}
\end{figure}

In order to explore further the relationships between AI usage and its perceived benefits, Spearman rank-order correlations were calculated using the mapped ordinal values from Q4 to Q7, see Figure \ref{fig:correlation-q4-q7}. The only  strong, statistically significant relationship  was found for \mbox{\emph{AI Usage \textrightarrow{} Faster Solving}} ($\rho = 0.64$, $p = 0.002$), indicating that students who reported higher levels of AI usage also tended to perceive faster task completion.

Taken together, these results highlight a general appreciation of AI tools as enablers of efficiency (coding speed-ups), with mixed perceptions of their value for concept comprehension and exploratory thinking. The latter exhibited weaker, more variable relationships that may warrant further investigation with a larger cohort.

\subsection{Qualitative Analysis of Open-Ended Questions}

To complement the quantitative findings, students’ answers to the open-ended survey questions were analysed using inductive content analysis. Responses were reviewed and coded to identify recurring themes and patterns, resulting in a set of emergent thematic categories per question. Each category is described with a brief interpretation, the number of student responses associated with it, and illustrative quotes. This qualitative layer offers detailed insight into students’ attitudes, experiences, and challenges when using AI code assistants during programming tasks. For the purposes of the analysis presented in the following sections, the question statements correspond to those included in Appendix~\ref{appendix:survey}. Representative responses are provided in English translation; the original Spanish versions are available in Appendix~\ref{appendix:responses}.

\paragraph{\textbf{Perceptions of programming learning}\\[.3cm]}

This subsection presents students’ reflections on how the use of AI assistants influenced their programming learning during the exam. The open-ended questions Q8-Q16 explored areas such as code understanding, debugging, design support, and originality. Responses provide insight into how students interacted with AI tools and how those interactions shaped their problem-solving approaches and learning outcomes.

\begin{table}[t]
\centering
\scriptsize
\renewcommand{\arraystretch}{1.4}
\setlength{\tabcolsep}{6pt}
\caption{Thematic categories for Q8 – Understanding AI-generated code}
\begin{tabular}{|p{2cm}|p{4cm}|c|p{5cm}|}
\hline
\textbf{Category} & \textbf{Explanation} & \textbf{$n$} & \textbf{Representative Quote (English)} \\
\hline
Clear \mbox{Understanding} & Students felt confident interpreting the AI-generated code due to prior mastery of syntax or concepts. & 7 & “Yes, because I understood all the basic programming being applied; Java syntax confused me a bit, but not much since I already know C++, which is very similar.” \\
\hline
Partial or \mbox{Conditional} Understanding & Students understood most of the code but noted confusion with advanced syntax or unfamiliar constructs. & 3 & “I understood a large part of the code, but there were some parts of the syntax that were difficult to grasp.” \\
\hline
Use of AI to \mbox{Request} \mbox{Clarifications} & Students proactively asked the AI assistant to explain confusing segments or debug errors. & 3 & “Yes, since I asked it to explain the code it generated.” \\
\hline
AI Output \mbox{Required} \mbox{Adaptation} & Students noted that although the output was useful, they had to adjust or rewrite it to match their understanding. & 3 & “By reviewing it a bit, almost everything could be understood… but it wasn’t exactly how we saw it in class.” \\
\hline
Minimal \mbox{Comprehension} Focus & Some students prioritised functional outcomes over fully understanding the code. & 1 & “Not really, because the goal was to get the program working, I didn’t stop to understand it piece by piece.” \\
\hline
\end{tabular}
\label{tab:q8-summary}
\end{table}

\paragraph{Understanding AI-generated code (Q8)\\[.3cm]}

Table~\ref{tab:q8-summary} presents the thematic classification of student responses regarding their understanding of AI-generated code during the first part of the exam. Most students reported clear comprehension ($n=7$), often due to prior knowledge of Java or similar languages. Others expressed partial understanding ($n=3$), frequently hindered by unfamiliar syntax. A notable subset actively sought clarification from the assistant ($n=3$), treating it as a coding mate. Several students ($n=3$) mentioned adapting or rewriting AI suggestions to align with class expectations or their personal style. Lastly, a minority ($n=1$) admitted deprioritizing comprehension in favor of getting a working solution. These results highlight that while AI code generation generally aligns with students’ understanding, deeper learning may still depend on intentional engagement and prior familiarity with the concepts or language.

\paragraph{Code adaptation (Q9)\\[.3cm]}

Table~\ref{tab:q9-main} shows that most students actively modified the AI-generated code, although the nature and depth of these changes varied considerably. A prevalent theme was the refinement of variable names, syntax, or code structure to enhance clarity, functionality, or alignment with the task’s requirements ($n = 10$). \mbox{Others} ($n = 6$) simplified or streamlined unnecessarily complex AI output. Some students went further, integrating new elements or refactoring logic to better suit their own problem-solving strategies ($n = 5$). A smaller group used the AI mainly for debugging or error-correction, without relying on it to produce full solutions ($n = 4$). Lastly, a few students did not engage in adaptation because they either generated the code themselves or only used AI to import libraries ($n = 3$). These findings reveal a broad spectrum of AI code usage: from passively receiving and executing suggestions to deeply engaging in rewriting, optimizing, and integrating ideas from the assistant.

\begin{table}[t]
\centering
\scriptsize
\renewcommand{\arraystretch}{1.4}
\setlength{\tabcolsep}{6pt}
\caption{Thematic categories for Q9 – Code adaptation}
\label{tab:q9-main}
\begin{tabular}{|p{2cm}|p{4cm}|c|p{5cm}|}
\hline
\textbf{Theme} & \textbf{Explanation} & \textbf{$n$} & \textbf{Representative Quote (EN)} \\
\hline
Refining and restructuring & Student changed variable names, method structure, or comments to clarify or better fit the problem & 10 & “I changed variable names and texts that would later appear on screen.” \\
\hline
Simplifying AI solutions & Student condensed or optimized overly verbose or complex AI-generated code & 6 & “Sometimes the AI gave long solutions and I realized I could make them shorter, so I modified them.” \\
\hline
Extending or improving AI output & Student added new elements, fixed logic, or repurposed suggestions to better match the task & 5 & “I added images of my choice and new options to what the AI gave me.” \\
\hline
Error-driven adaptation & Student primarily used the AI to identify or solve code errors, adjusting code accordingly & 4 & “I used the AI to correct errors and learn more about object-oriented programming.” \\
\hline
Minimal or no adaptation & Student used AI minimally or only for basic components like libraries, without deeper changes & 3 & “No, I only asked it to show me the errors and give me the libraries.” \\
\hline
\end{tabular}
\end{table}

\paragraph{Confidence in AI-generated code (Q10)\\[.3cm]}

Table~\ref{tab:q10} presents the categories that emerged when students described their confidence using the code generated by the AI. A majority of students ($n=7$) reported feeling confident with the AI-generated code, citing ease of use, clarity, and correctness. Others ($n=6$)  expressed partial confidence, noting that while the code worked, it sometimes introduced inefficiencies or concepts they didn't fully understand. A few students ($n=6$) felt little or no confidence, pointing to previous experiences with errors or a lack of understanding of how the code worked.

\begin{table}[t]
\centering
\scriptsize
\renewcommand{\arraystretch}{1.4}
\setlength{\tabcolsep}{6pt}
\caption{Thematic categories for Q10 – Confidence in AI-generated code}
\label{tab:q10}
\begin{tabular}{|p{2cm}|p{4cm}|c|p{5cm}|}

\hline
\textbf{Category} & \textbf{Explanation} & \textbf{$n$} & \textbf{Representative Quote} \\
\hline
High confidence & Students felt confident with the code provided by the AI and often mentioned its usefulness or clarity. & 7 & “I felt confident because it solves problems very easily and I had clarity about my code.” \\
\hline
Moderate \mbox{confidence} with doubts & Students were generally confident but expressed concerns about efficiency, correctness, or unfamiliar code structures. & 6 & “At first I felt confident using the AI-generated code, but I had some doubts—whether it was efficient or had hidden errors.” \\
\hline
Partial \mbox{confidence} or skepticism & Students used the code with caution, aware that it might be incorrect or not optimal. & 4 & “Sometimes I used the code and it worked perfectly, but other times it generated errors or used unfamiliar concepts I didn’t understand.” \\
\hline
Low confidence or distrust & Students distrusted the AI-generated code, often due to prior errors or lack of understanding. & 2 & “I didn’t feel confident, I always stopped to check the code because the AI often gave me wrong solutions.” \\
\hline
\end{tabular}
\end{table}

\paragraph{Able to solve without AI (Q11) \& Second part easier than first (Q12)\\[.3cm]}

\begin{figure}[t]
\centering
\includegraphics[width=.8\textwidth]{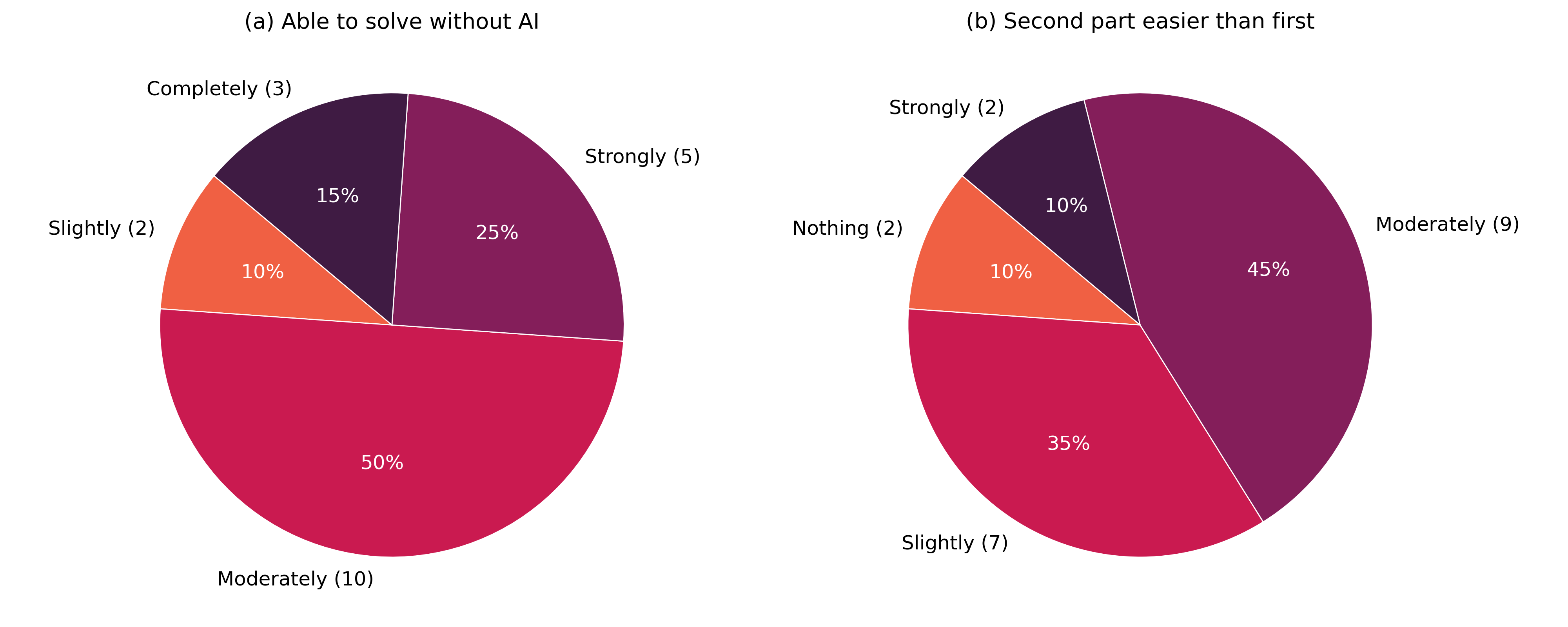}
\caption{Student responses to Q11 and Q12. (a) Self-reported ability to solve the exercise without AI. (b) Perceived difficulty of the second part relative to the AI-assisted first part.}
\label{fig:pies-q11-q12}
\end{figure}

Unlike the open-ended responses analyzed earlier, these two items are Likert-type and categorical, respectively. We therefore applied a descriptive approach similar to that used for Q4--Q7. Figure~\ref{fig:pies-q11-q12} summarizes students’ self-reported ability to complete the second-part exercise without AI (Q11) and their perception of its difficulty compared to the earlier, AI-assisted task (Q12). In both cases, responses concentrate around the middle of the scale: over half of the cohort selected \textit{Moderately} and another 10\% \textit{Slightly}, suggesting that many students did not feel entirely confident solving the task unaided. Only three respondents (15\%) chose \textit{Completely}, indicating full confidence in their independent performance. Likewise, judgments of relative difficulty skew toward the lower end: the majority reported the unaided part was only \textit{Moderate}, \textit{Slightly} or \textit{Nothing} easier, if at all.

\begin{figure}[t!]
  \centering
  \includegraphics[width=.6\textwidth]{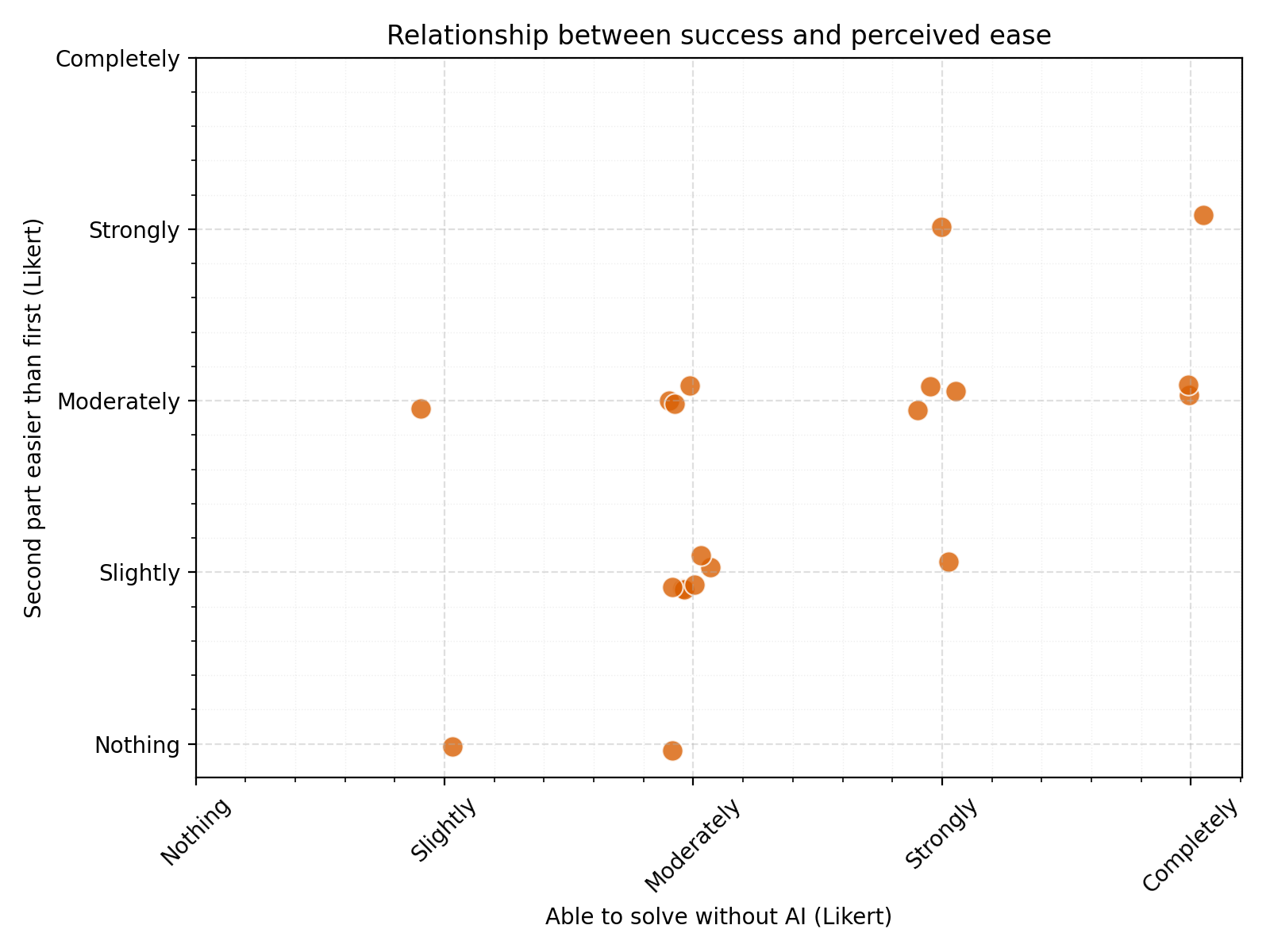}
  \caption{Correlation between perceived unaided ability (Q11) and relative ease of second part (Q12).}
  \label{fig:scatter-q11-q12}
\end{figure}

To further explore this relationship, Figure~\ref{fig:scatter-q11-q12} presents a scatter plot of the two items (with jitter added for clarity). The trend is positive and statistically significant. Students who felt more capable of solving the task without AI also tended to judge the second part as easier. Spearman correlation confirms a moderate-to-strong association ($\rho = .58$, $p = .008$), with approximately 34\% of the variance in perceived ease explained by students’ self-reported ability to complete the task independently. This pattern underscores how perceived difficulty is shaped by actual performance: those who succeeded unaided experienced less cognitive strain, whereas those who struggled judged the section more demanding.

\paragraph{Transfer to second part (Q13)\\[.3cm]}

Table~\ref{tab:q13} presents the themes that emerged from student responses to whether using AI in the first part helped with the second part of the exam. Most students ($n=6$) indicated that the AI provided a foundational understanding of the code and its structure, which they could leverage in the second part ($n=4$). This included understanding GUI elements, identifying modifiable sections, and reusing functional components. Some ($n=3$) emphasized how AI-assisted debugging in the first part made them more prepared to correct mistakes independently later. However, a few students ($n=3$) stated that the AI's advanced output created dependency or confusion, making the second part more difficult. Others ($n=4$) mentioned that their minimal or lack of use of AI in the first part meant no noticeable benefit in the second.

\begin{table}[t!]
\centering
\scriptsize
\renewcommand{\arraystretch}{1.4}
\setlength{\tabcolsep}{6pt}
\caption{Thematic categories for Q13 – Transfer to second part}
\label{tab:q13}
\begin{tabular}{|p{2cm}|p{4cm}|c|p{5cm}|}
\hline
\textbf{Category} & \textbf{Explanation} & \textbf{$n$} & \textbf{Representative Quote} \\
\hline
Conceptual scaffolding & AI support in the first part helped students better understand code logic, structures, or syntax, aiding their performance later. & 6 & “Using AI gave me a foundation so I could later make modifications to the code by myself.” \\
\hline
Code reusability & Students reused or adapted code or structures generated with AI in the first part for solving problems in the second part. & 4 & “There were things in the code I could reuse to solve the new problems.” \\
\hline
Debugging \mbox{preparedness} & Experience with AI-assisted debugging helped students anticipate or fix similar issues in the second part. & 3 & “It helped me a bit to prepare for possible errors in the program later on.” \\
\hline
Partial \mbox{understanding} & Students noted some help from AI, though limited by gaps in theoretical or syntactic knowledge. & 3 & “It was a bit easier to declare functions and constructors as the AI had suggested, but even so... it was a bit complicated.” \\
\hline
No benefit & Students did not perceive any advantage from using AI in the second part, either due to not using it earlier or due to confusion created by overreliance. & 4 & “I think the AI actually made the first part too easy, which made the second part a real challenge.” \\
\hline
\end{tabular}
\end{table}

\paragraph{Confidence without AI (Q14)\\[.3cm]}

Table~\ref{tab:q14} shows the thematic categories that emerged when students reflected on their confidence while solving the second part of the exam without AI assistance. Many students ($n=12$) expressed high or moderate confidence, often attributing it to their understanding of the code, prior preparation, or familiarity with the structure. However, several students ($n=6$) expressed low confidence, citing difficulties with understanding Java syntax, feeling uncertain about their own code, or missing the reassurance provided by the AI. A few($n=2$) described mixed or evolving feelings of confidence as they progressed through the task.

\begin{table}[t!]
\centering
\scriptsize
\renewcommand{\arraystretch}{1.4}
\setlength{\tabcolsep}{6pt}
\caption{Thematic categories for Q14 – Confidence without AI}
\label{tab:q14}
\begin{tabular}{|p{2cm}|p{4cm}|c|p{5cm}|}
\hline
\textbf{Category} & \textbf{Explanation} & \textbf{$n$} & \textbf{Representative Quote} \\
\hline
High confidence & Students felt comfortable or confident solving the task without AI support, based on their preparation or understanding. & 7 & “Yes, because I understood the concepts and had a mental image of how to do it.” \\
\hline
Moderate confidence & Students expressed some confidence, though sometimes with uncertainty or initial doubts. & 5 & “More or less, I felt comfortable because I thought I knew how my program worked.” \\
\hline
Low confidence & Students reported lack of confidence due to difficulty with syntax, logic, or unfamiliar code. & 6 & “No, I don’t have the tools to modify it—I don’t know Java.” \\
\hline
Mixed or partial confidence & Students described changing feelings—initial uncertainty that shifted toward confidence, or cases where understanding was partial. & 2 & “A bit, because I knew what the code did, but there were things I didn’t know how they worked.” \\
\hline
\end{tabular}
\end{table}

\paragraph{Difficulties without AI (Q15)\\[.3cm]}

Table~\ref{tab:q15} presents the difficulties students faced in the second part of the exam when they no longer had access to the AI. The most frequent challenge was related to syntax and language-specific knowledge, particularly Java. Students ($n=6$) mentioned issues such as semicolons, method declarations, and array handling. Others ($n=4$) struggled with debugging or lacked confidence without immediate feedback or guidance. Some students ($n=3$) experienced conceptual difficulties, realizing gaps in their understanding once they had to code independently. A few students ($n=3$) reported difficulties in devising the solution to task-specific challenges, whether others ($n=3$) had no major difficulties, citing simplicity of tasks or sufficient preparation.

\begin{table}[t]
\centering
\scriptsize
\renewcommand{\arraystretch}{1.4}
\setlength{\tabcolsep}{6pt}
\caption{Thematic categories for Q15 – Difficulties without AI support}
\label{tab:q15}
\begin{tabular}{|p{2cm}|p{4cm}|c|p{5cm}|}
\hline
\textbf{Category} & \textbf{Explanation} & \textbf{$n$} & \textbf{Representative Quote} \\
\hline
Syntax and language-specific issues & Students struggled with Java syntax, unfamiliar functions, or method usage. & 6 & “I think understanding Java functions and the correct syntax was my biggest difficulty.” \\
\hline
Debugging and error-handling & Students missed the instant support for checking errors or confirming their approach. & 4 & “The biggest issue was not having a reference to check if I was on the right track, especially when I had doubts about the code.” \\
\hline
Conceptual gaps & Some students realized they did not fully understand basic concepts when coding independently. & 3 & “You think you understood how the code worked, but trying to create it on your own shows you didn’t fully get it.” \\
\hline
Task-specific challenges & Students mentioned difficulties tied to specific functions or requirements in the task. & 3 & “I found the part about restoring the billboard, especially the backup part, a bit difficult.” \\
\hline
No major \mbox{difficulty} & Students reported no significant issues, citing time or general simplicity as the only challenge. & 3 & “None, just the time it took to do it.” \\
\hline
\end{tabular}
\end{table}

\paragraph{Impact of AI on second part (Q16)\\[.3cm]}

Table~\ref{tab:q16} summarises student responses on how their performance might have changed if they had access to AI in the second part of the exam. Most students ($n=9$) stated that AI would have improved their efficiency, helping them complete the tasks faster and with fewer errors. Some ($n=3$) mentioned that AI would have increased their confidence, especially during debugging or identifying syntax issues. A few students ($n=3$) acknowledged a trade-off: although AI would have helped with speed, it might have reduced their learning or understanding of the code. On the other hand, several students ($n=4$) reported that AI would have made little or no difference to their final performance.

\begin{table}[t]
\centering
\scriptsize
\renewcommand{\arraystretch}{1.4}
\setlength{\tabcolsep}{6pt}
\caption{Thematic categories for Q16 – Impact of AI on second part}
\label{tab:q16}
\begin{tabular}{|p{2cm}|p{4cm}|c|p{5cm}|}
\hline
\textbf{Category} & \textbf{Explanation} & \textbf{$n$} & \textbf{Representative Quote} \\
\hline
Increased \mbox{efficiency} and speed & AI would have helped students complete tasks faster, reducing time spent debugging or writing code. & 9 & “It would’ve been much easier because we would’ve saved time figuring out where and how to do something.” \\
\hline
Confidence and error reduction & Students believe AI would have improved their accuracy or helped fix errors. & 3 & “It would have helped me debug errors and optimize the code, improving my performance.” \\
\hline
Trade-off: speed vs. learning & Students recognized that AI might improve performance but reduce learning opportunities. & 3 & “AI would have helped with time, but delays the development of problem-solving and logic skills.” \\
\hline
No significant impact & Students felt AI access would not have affected their performance significantly. & 4 & “It would’ve been exactly the same.” \\
\hline
\end{tabular}
\end{table}

\paragraph{\textbf{Perceptions of the exam format and experience}\\[.3cm]}

This subsection focuses on students’ perceptions of the exam format itself, independent of specific programming content. It addresses how they experienced the two-part structure, the use of AI in the first phase, time constraints, grading perceptions, and overall satisfaction. These reflections are relevant for evaluating the exam as a pedagogical intervention and for improving future implementations.

\paragraph{General perceptions of the exam format (Q17)\\[.3cm]}

Table~\ref{tab:q17} summarizes students’ overall perceptions of the exam format. Many students ($n=6$) described the experience as well-structured or enjoyable, highlighting the novelty of the format and its focus on code understanding rather than just code creation. Several responses ($n=5$) emphasized the exam’s usefulness for reinforcing learning and applying concepts, despite challenges. A subset of students ($n=4$) acknowledged the exam was challenging or stressful, particularly during the second part without AI. Some ($n=3$) described the experience as neutral or uneventful. A final group ($n=4$) offered constructive suggestions to improve pacing, session structure, or task balance, such as preferring the exam to be done in one session or adjusting the time given for specific tasks.

\begin{table}[t]
\centering
\scriptsize
\renewcommand{\arraystretch}{1.4}
\setlength{\tabcolsep}{6pt}
\caption{Thematic categories for Q17 – General perceptions of the exam format}
\label{tab:q17}
\begin{tabular}{|p{2cm}|p{4cm}|c|p{5cm}|}
\hline
\textbf{Category} & \textbf{Explanation} & \textbf{$n$} & \textbf{Representative Quote} \\
\hline
Positive and well-structured & Students found the format interesting, novel, or effective in assessing applied knowledge. & 6 & “I liked the format... it tests our ability to modify a code that could have been made 100\% by AI.” \\
\hline
Reinforced learning & Students found the exam useful for understanding code, applying concepts, or engaging with the content. & 5 & “It was a challenging but useful experience... it was a good opportunity to apply what I had learned.” \\
\hline
Stressful or \mbox{demanding} & Some students found the second part more stressful, citing uncertainty or time pressure. & 4 & “The second part was a bit more stressful since we had to modify the code just by looking at it.” \\
\hline
Neutral & Some students gave short, non-committal evaluations, suggesting a neutral perception. & 3 & “Good, I had a good feeling about the exam.” \\
\hline
Suggestions for improvement & Students suggested changes to timing, structure, or balance between tasks. & 4 & “It’s preferable to finish the exam in a single session, as usual.” \\
\hline
\end{tabular}
\end{table}

\paragraph{Comparison with and without AI (Q18)\\[.3cm]}

Table~\ref{tab:q18} summarizes student reflections after comparing their performance with and without AI assitance. A large group ($n=9$) emphasized that AI increased their efficiency and confidence, helping them resolve problems more quickly and with less uncertainty. Several students ($n=4$) acknowledged that although AI was useful, it could also hinder deep understanding or independent learning. A few ($n=3$) focused on how the experience revealed specific areas for improvement in their programming skills. Other students ($n=2$) saw the exercise as a learning experience, noting that the AI-aided part better prepared them for working independently. Lastly, two students ($n=2$) gave neutral or unclear responses, with no conclusive insight drawn from the comparison.

\begin{table}[t]
\centering
\scriptsize
\renewcommand{\arraystretch}{1.4}
\setlength{\tabcolsep}{6pt}
\caption{Thematic categories for Q18 – Comparison with and without AI}
\label{tab:q18}
\begin{tabular}{|p{2cm}|p{4cm}|c|p{5cm}|}
\hline
\textbf{Category} & \textbf{Explanation} & \textbf{$n$} & \textbf{Representative Quote} \\
\hline
Increased \mbox{efficiency} and confidence with AI & AI helped students resolve problems faster and feel more confident. & 9 & “With AI I was able to move faster and feel more confident... it made the process more efficient.” \\
\hline
AI hinders deep understanding & While helpful, AI can give a false sense of understanding or hinder long-term learning. & 4 & “AI helps a lot, but doesn’t give enough time to truly understand the concepts.” \\
\hline
Revealed areas for improvement & The comparison helped students identify gaps in logic, syntax, or organization. & 3 & “I need to better organize my code and practice more to be faster and more efficient.” \\
\hline
AI as preparation for autonomy & Using AI in the first part helped students feel ready to work alone in the second. & 2 & “In the part with AI I resolved doubts that prepared me for the second part.” \\
\hline
Neutral or no clear conclusion & Students did not express a clear insight or conclusion from the comparison. & 2 & “None, I don’t have any conclusion.” \\
\hline
\end{tabular}
\end{table}

\paragraph{Exam really assessed programming skills (Q19)\\[.3cm]}

Table~\ref{tab:q19} presents students’ reflections on whether the exam effectively assessed their programming knowledge and skills. Most respondents ($n=15$) believed the exam was a fair and comprehensive assessment, citing the need to apply core concepts (including object-oriented programming) and reason through code with and without AI assistance. A smaller group ($n=3$) expressed partial agreement, suggesting that while the exam did test knowledge, it didn’t fully capture their skills or had uneven emphasis across sections. Only one student ($n=1$) felt the exam did not adequately assess their programming abilities, noting it focused more on prompt crafting and general logic than on technical depth.

\begin{table}[t]
\centering
\scriptsize
\renewcommand{\arraystretch}{1.4}
\setlength{\tabcolsep}{6pt}
\caption{Thematic categories for Q19 – Exam really assessed programming skills}
\label{tab:q19}
\begin{tabular}{|p{2cm}|p{4cm}|c|p{5cm}|}
\hline
\textbf{Category} & \textbf{Explanation} & \textbf{$n$} & \textbf{Representative Quote} \\
\hline
Yes, comprehensive assessment & Students felt the exam required them to apply key programming concepts and demonstrate their understanding with and without AI. & 15 & “Although AI helped in some parts, the exam made me think and use my own knowledge to solve programming problems.” \\
\hline
Partially \mbox{assessed} & The exam tested some aspects of knowledge but not others, or lacked balance between parts. & 3 & “It assessed my knowledge, but I feel it didn’t evaluate my skills as much.” \\
\hline
No, limited \mbox{assessment} & One student argued that the exam focused more on prompting and general logic than on technical programming skills. & 1 & “No, only on prompt writing and general programming logic.” \\
\hline
\end{tabular}
\end{table}

\paragraph{Suggestions to improve the exam (Q20)\\[.3cm]}

Table~\ref{tab:q20} summarizes students' suggestions to improve the exam format. The most common recommendation ($n=5$) was to adjust the timing—either by shortening the exam duration, conducting it in one class, or reducing the number of sessions involved. A second relevant group ($n=4$) suggested providing better guidance, such as clearer instructions, structured code templates, or examples. Others recommended enhancing the design or focus of the exam ($n=4$), particularly regarding how AI is used and the balance between prompting and programming. A few students ($n=3$) emphasized better preparation, especially more practice in exam-like conditions or earlier exposure to the programming language. Interestingly enough, some ($n=4$) students said that they had no suggestions. Finally, issues related to fairness or inconsistency in the grading were marginally mentioned ($n=2$).

\begin{table}[t]
\centering
\scriptsize
\renewcommand{\arraystretch}{1.4}
\setlength{\tabcolsep}{5pt}
\caption{Thematic categories for Q20 – Suggestions to improve the exam}
\label{tab:q20}
\begin{tabular}{|p{2cm}|p{4cm}|c|p{5cm}|}
\hline
\textbf{Category} & \textbf{Explanation} & \textbf{$n$} & \textbf{Representative Quote} \\
\hline
Adjust timing & Suggestions to shorten the duration or number of sessions, or to streamline the exam process. & 5 & “That it should be done in a single session.” \\
\hline
Better guidance & Requests for clearer instructions, structured code, or examples. & 4 & “Provide greater clarity on the requirements of each exercise.” \\
\hline
Improve exam \mbox{design} & Suggestions to refine how AI is used, increase focus, or make the exam more concise. & 4 & “That the first part should include AI, but in a more restricted way.” \\
\hline
More preparation or practice & Emphasis on having more exam-like practice or earlier exposure to the programming language. & 3 & “More prior familiarization with the language that will be used.” \\
\hline
No suggestions & Students explicitly said they had no suggestions. & 4 & “None.” \\
\hline
Fairness or grading issues & Concerns about uneven workload or evaluation inconsistency. & 2 & “Avoid too many extra questions [...] students who were graded first had fewer methods.” \\
\hline
\end{tabular}
\end{table}

\paragraph{AI useful for learning programming (Q21) \& AI replaces coding skills (Q22) \\[.3cm]}

\begin{figure}[t]
\centering
\includegraphics[width=.8\textwidth]{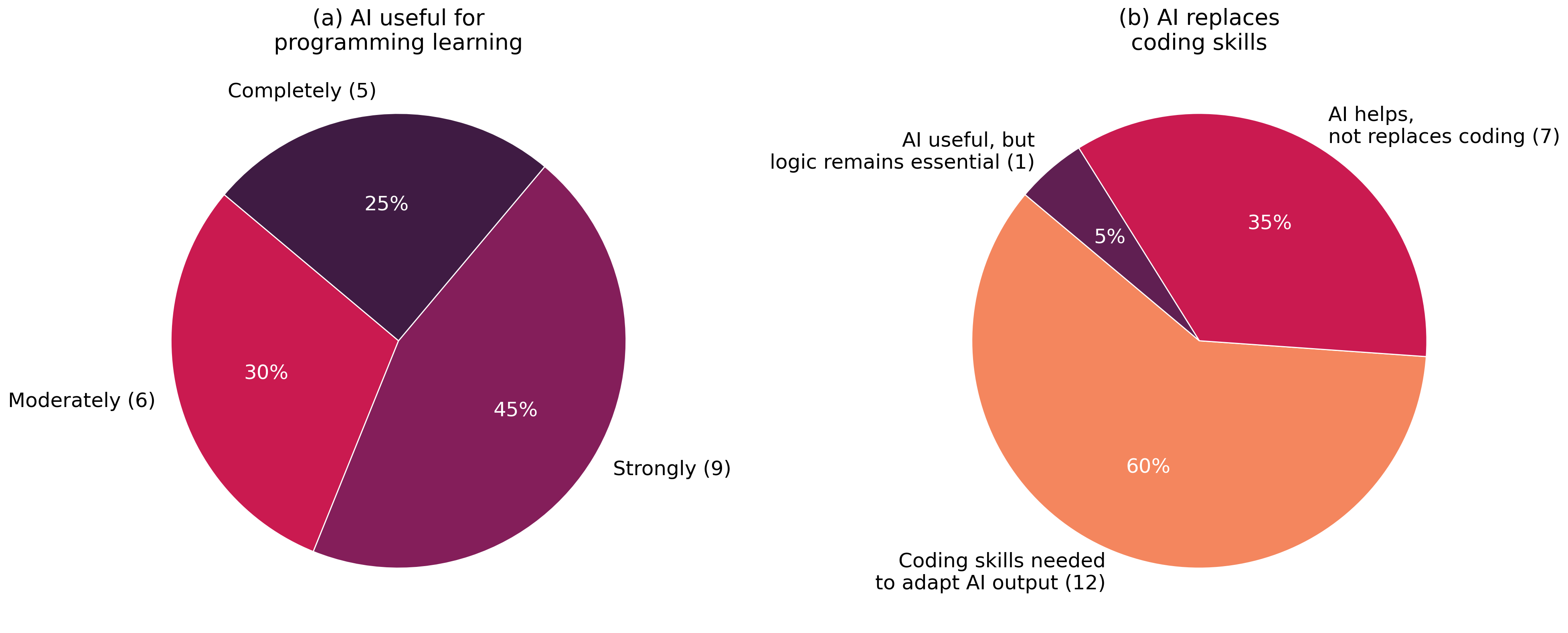}
\caption{Student responses to Q21 and Q22. (a) Usefulness of AI for programming learning. (b) Perceived need for mastering coding skills alongside AI.}
\label{fig:q21-q22}
\end{figure}

In contrast to the open-ended questions analyzed in this subsection, Q21 and Q22 used Likert-scale and categorical formats, respectively, so we performed a descriptive analysis as we did for Q4--Q7. Figure~\ref{fig:q21-q22} illustrates students' perceptions of the role AI plays in programming education. Panel~(a) reflects a strong consensus that AI is useful for learning programming: three-quarters selected \textit{Strongly} or \textit{Completely}, while the rest responded \textit{Moderately}. No student chose \textit{Nothing} or \textit{Slightly}, yielding a high overall mean ($\approx 3.95$ on a 1--5 scale). Panel~(b) further clarifies students’ views on AI in relation to acquiring core coding skills. A majority ($n=12$) agreed that \textit{Coding skills needed to adapt AI output} emphasizing the importance of understanding logic and syntax. Others ($n=7$) selected \textit{AI helps, not replaces coding} reinforcing the idea of AI as a support tool rather than a substitute. One respondent chose \textit{AI useful, but logic remains essential} echoing a similar sentiment. Overall, the charts depict an enthusiastic yet pragmatic cohort: students appreciate AI’s role in supporting learning, while remaining committed to acquiring the foundational programming skills necessary to use it effectively.

\paragraph{Future use of AI in programming (Q23)\\[.3cm]}

Table~\ref{tab:q23} summarizes students' expectations and preferences regarding the future use of AI in their learning and programming processes. The most common intention ($n=6$) was to use AI as a \textit{support for learning and debugging}, such as solving doubts, understanding concepts, or reviewing syntax. A second group ($n=5$) emphasized a \textit{selective or cautious use}, intending to rely on AI for reviewing or optimizing code without replacing their own understanding. Another frequent category ($n=4$) was the use of AI for \textit{automation and optimization}, especially in repetitive or time-consuming tasks. Some students ($n=3$) preferred \textit{minimal or no use}, motivated by the desire to build skills independently. Finally, a smaller group ($n=2$) highlighted their interest in \textit{learning from AI explanations}, such as asking AI to teach programming concepts or explain documentation.

\begin{table}[t]
\centering
\scriptsize
\renewcommand{\arraystretch}{1.4}
\setlength{\tabcolsep}{5pt}
\caption{Thematic categories for Q23 – Future use of AI in programming}
\label{tab:q23}
\begin{tabular}{|p{2cm}|p{4cm}|c|p{5cm}|}
\hline
\textbf{Category} & \textbf{Explanation} & \textbf{$n$} & \textbf{Representative Quote} \\
\hline
Support for learning and debugging & Using AI as an aid to resolve doubts, understand code, or support learning processes. & 6 & “I would like to use AI as a support tool to resolve doubts, debug code, and get suggestions.” \\
\hline
Selective or cautious use & Using AI for specific tasks (e.g., code review, error detection) while maintaining control of the learning or coding process. & 5 & “To solve problems that are difficult for me, but not to do everything for me.” \\
\hline
Automation and optimization & Using AI to speed up repetitive tasks or optimize existing code. & 4 & “To optimize time in large projects, letting AI handle the simplest parts.” \\
\hline
Minimal or no use & Preference for learning without AI or using it only as a last resort, to ensure personal mastery. & 3 & “I would rather learn entirely on my own so that in the end it’s just another tool.”  \\
\hline
Learning from AI explanations & Asking AI to teach programming topics, summarize documentation, or explain solutions. & 2 & “Asking it to explain processes or summarize documentation.” \\
\hline
\end{tabular}
\end{table}

\paragraph{AI role in programming courses (Q24)\\[.3cm]}

Table~\ref{tab:q24} summarizes students’ perspectives on how AI should be integrated into programming courses. The most frequent category ($n=6$) was to use AI as a \textit{support tool}, especially for solving doubts, debugging, or accelerating coding processes. A second group ($n=5$) preferred a \textit{cautious or selective use}, advocating for strong conceptual foundations before AI involvement, or emphasizing autonomous use outside core teaching. Another notable group ($n=4$) explicitly stated a preference for \textit{no use of AI} in coursework, emphasizing the value of pedagogy and human instruction. Finally, a smaller group ($n=2$) viewed AI integration as a chance to \textit{learn about AI systems themselves} or incorporate them as course content.

\begin{table}[t]
\centering
\scriptsize
\renewcommand{\arraystretch}{1.4}
\setlength{\tabcolsep}{5pt}
\caption{Thematic categories for Q24 – AI role in programming courses}
\label{tab:q24}
\begin{tabular}{|p{2cm}|p{4cm}|c|p{5cm}|}
\hline
\textbf{Category} & \textbf{Explanation} & \textbf{$n$} & \textbf{Representative Quote} \\
\hline
Support tool & Using AI as a classroom aid for solving doubts, debugging, or generating examples. & 6 & “I would like AI to be used in programming courses to help us solve doubts, debug code, and give clearer explanations when something is not understood.” \\
\hline
Autonomous use & Prioritizing traditional teaching and introducing AI only for autonomous learning or secondary use. & 5 & “It should be part of autonomous learning, but not a tool for teaching; it’s easier to understand concepts explained by a professor.” \\
\hline
No AI in courses & Rejecting AI integration in teaching, favoring traditional pedagogy and deep understanding. & 4 & “Not at all, maintaining good pedagogy and fully covering the concepts; later the student will decide how to use AI, but with the knowledge already acquired.” \\
\hline
Understanding AI systems & Interest in learning how AI works or integrating it into course material itself. & 2 & “I would like to learn how they work and create one.” \\
\hline
\end{tabular}
\end{table}

\section{Discussion}

This small-scale study offers preliminary insights into the role of generative AI assistants in programming education. Our combined survey analysis and in-person observations reveal a complex pattern: while students perceived AI as helpful—especially for syntactic and structural guidance—this support appeared largely superficial. When the tools were removed, many students struggled to apply concepts independently, suggesting a limited transfer of knowledge. This pattern points to a possible \emph{cognitive offloading} effect, where a reliance on AI may have replaced, rather than reinforced, core problem-solving processes.

Further evidence of this limited understanding emerged from students’ self-reported confidence, which dropped significantly during the unaided session. Despite initial optimism, students frequently expressed uncertainty with debugging and logic formulation, highlighting a gap in their internalized knowledge even when their AI-supported performance seemed strong.

These findings also underscore a deeper issue: the disconnect between self-perception and observed behavior. For example, while many students rated their AI usage as Moderate, instructor observations indicated widespread and enthusiastic reliance. Similarly, students described the second task as Moderately difficult, yet visibly struggled with it despite its simpler structure. These contrasts suggest that learners may be unaware of the extent of their reliance on AI.

Interestingly, students overwhelmingly valued the exam format itself. Many described the two-session design as fair and educational, helping them directly recognize both the benefits and limitations of AI assistance. They also highlighted the need to use AI tools in a way that supports—not replaces—their own reasoning, favoring a balanced approach that encourages active engagement and skill development.

Overall, these results underscore the need for thoughtful AI integration in programming education. Scaffolding activities that require students to adapt or critique AI-generated code may help mitigate overreliance and support the development of core skills. We remark that despite being limited by its small sample size and specific context, this study's findings provide a foundation for future research that incorporates behavioral, longitudinal, and performance metrics to better understand the long-term impact of AI-assisted learning.

\section{Conclusions}
The rapid adoption of generative AI in programming education raises critical questions about its long-term impact on learning. In this context, this exploratory study suggests that while students perceive AI support as valuable, their independent performance may not align with their sense of mastery, indicating a possible overestimation of their own skills. This disconnect highlights the potential for cognitive offloading, where reliance on the tool replaces the internalization of knowledge. These findings underscore the importance of designing activities that use AI for scaffolding, not to replace the essential practice of independent problem-solving. Ultimately, our results present a compelling case for reimagining programming education to foster active engagement and strengthen computational thinking.


\appendix

\section*{Appendixes}

\section{Exam Problem Statement}
\label{app:exam-problem}

\noindent\textbf{Movie Billboard}  
\medskip

The goal of this exam is to assess the understanding and application of Object-Oriented Programming (OOP) principles, including inheritance, basic graphical interfaces with dialog windows, and the management of object collections. You are required to develop a simple Java program to manage a movie theater’s billboard. The program should allow for the following functionalities:

\begin{enumerate}
    \item \textbf{Automatic creation of an initial billboard:}  
    Upon starting the program, a predefined list of movies currently showing must be automatically generated, including their details (title, genre, duration, and showtimes).
    \medskip

    \item \textbf{Viewing the billboard:}  
    Through dialog windows, the user should be able to view the list of movies along with their details and showtimes.
    \medskip

    \item \textbf{Adding new movies:}  
    The program should allow the user to input information for new movies (title, genre, duration, and showtimes) through dialog windows and add them to the billboard.
\end{enumerate}

\noindent \textbf{Specific Requirements:}
\begin{itemize}
    \item[\textbullet] \textbf{Use of Inheritance:}  
    A class hierarchy using inheritance must be designed to represent different aspects of movies.
    \medskip

    \item[\textbullet] \textbf{User Interface with Dialog Windows:}  
    Interaction with the user must be implemented using dialog windows (\texttt{JOptionPane}) to display information and collect input. You may use \texttt{JLabel} and \texttt{JPanel} to organize content within dialogs.
    \medskip

    \item[\textbullet] \textbf{Controller Class:}  
    A controller class should manage the business logic, user interaction, and the billboard’s data.
    \medskip

    \item[\textbullet] \textbf{Movie Collection:}  
    The billboard must be managed using a collection of objects (e.g., an \texttt{ArrayList}).
    \medskip

    \item[\textbullet] \textbf{Automatic Initialization:}  
    The initial list of movies should be hardcoded within the program (no need for file reading).
    \medskip

    \item[\textbullet] \textbf{Data Entry:}  
    New movies should be added using dialog windows that request the necessary information from the user.
\end{itemize}

\section{Extension Task Set}
\label{app:extensions}

\noindent In the second part of the exam, students were required to manually extend or modify their existing Java program without assistance from AI tools. The following short programming tasks were presented, each targeting specific concepts such as object manipulation, searching, collections, and interface logic:

\begin{enumerate}
    \item \textbf{Clear the billboard:} Implement a function to remove all movies from the current list.
    \medskip

    \item \textbf{Count movies only:} Count and display the number of items that are specifically movies, excluding other types of events (e.g., theater plays or sports matches).
    \medskip

    \item \textbf{Cancel a category (e.g., opera or format):} Remove all events belonging to a specific category—such as operas or 3D movies—from the billboard.
    \medskip

    \item \textbf{Football-only billboard} Reconfigure the billboard so that only football-related events (e.g., UEFA Champions League final) are scheduled in all available showtimes.
    \medskip

    \item \textbf{Undo billboard clearing:} Allow the restoration of the billboard content after a full clear operation has been performed.
    \medskip

    \item \textbf{Duplicate the billboard:} Create a copy of the current billboard entries and append them to the existing list.
    \medskip

    \item \textbf{Find the longest event:} Identify and display the event with the longest duration.
    \medskip

    \item \textbf{List short events} Display all events with a duration of less than 90 minutes.
    \medskip

    \item \textbf{Show events at 3:00 p.m.:} List all events scheduled to begin at 3:00 p.m.
    \medskip

    \item \textbf{Delete showtimes at 3:00 p.m. and 5:00 p.m.:} Remove these specific showtimes from all events on the billboard.
\end{enumerate}

\section{Survey Instrument}
\label{appendix:survey}

This appendix presents the full set of survey questions administered to participants in Session 3. The survey was designed to capture students' experiences, perceptions, and attitudes regarding the use of AI code assistants during the programming examination. The table below lists each item in its original Spanish formulation, alongside its English-coded version as used in this paper for analysis and reporting purposes, the type of question (e.g., Likert scale, multiple choice, open-ended), and a unique identifier for ease of reference. The survey was anonymous, collected no sensitive personal data, and thus did not require formal ethical approval.

\begin{table}[t]
\centering
\caption{Survey questions: original in Spanish, English label (shortened translation), and question types}

{\scriptsize
\setlength{\tabcolsep}{4pt}
\renewcommand{\arraystretch}{1.2}
\begin{tabularx}{\textwidth}{|c|p{5.5cm}|p{3.5cm}|X|}
\hline
\multicolumn{1}{|c|}{\textbf{ID}} &
\multicolumn{1}{c|}{\textbf{Original}} &
\multicolumn{1}{c|}{\textbf{English Label}} &
\multicolumn{1}{c|}{\textbf{Question}} \\
\multicolumn{1}{|c|}{} &
\multicolumn{1}{c|}{\textbf{(Spanish)}} &
\multicolumn{1}{c|}{\textbf{(short translation)}} &
\multicolumn{1}{c|}{\textbf{Type}} \\
\hline\hline

Q1 & Nivel de dominio en programación (antes del examen): & Programming proficiency  & Multiple choice \\
\hline
Q2 & Herramientas de IA utilizadas en la primera sesión del exámen: & AI-Code Assistant & Multiple \mbox{response} \\
\hline
Q3 & Tareas específicas con IA: & AI use cases & Multiple \mbox{response} \\
\hline
Q4 & Nivel de uso de IA durante el examen: & Extent of AI usage & Likert scale \\
\hline
Q5 & ¿En la primera parte del examen, la IA me ayudó a resolver el ejercicio más rápido? & AI helped solve exercise faster & Likert scale \\
\hline
Q6 & ¿En la primera parte del examen, la IA me ayudó a comprender mejor los conceptos? & AI helped understand concepts better & Likert scale \\
\hline
Q7 & ¿En la primera parte del examen, la IA me permitió explorar diferentes soluciones? & AI helped explore different solutions & Likert scale \\
\hline
Q8 & ¿En la primera parte del examen, entendías el código que generó la IA? ¿Por qué si o por qué no? & Understanding AI-generated code & Open-ended \\
\hline
Q9 & ¿En la primera parte del examen, modificaste o adaptaste el código generado por la IA? ¿Cómo? & Code adaptation & Open-ended \\
\hline
Q10 & ¿Te sentiste confiado al usar el código generado por la IA? ¿Quedaste con alguna inquietud? & Confidence in AI-generated code & Open-ended \\
\hline
Q11 & ¿En la segunda parte del examen, sin ayuda de la IA, pudiste resolver el ejercicio? & Able to solve without AI & Likert scale \\
\hline
Q12 & ¿En términos del nivel de dificultad y/o esfuerzo, consideras que la segunda parte del examen era más sencilla de resolver que la primera parte? & Second part easier than first & Likert scale \\
\hline
Q13 & ¿Crees que usar IA en la primera parte del examen te ayudó para la segunda parte? ¿Cómo? & Transfer to second part & Open-ended \\
\hline
Q14 & ¿Te sentiste cómodo o confiado resolviendo el ejercicio sin IA? ¿Por qué? & Confidence without AI & Open-ended \\
\hline
Q15 & ¿En la segunda parte sin acceso a la IA, encontraste alguna dificultad? ¿Cuál fue la mayor? & Difficulties without AI & Open-ended \\
\hline
Q16 & Si hubieras tenido acceso a IA en la segunda parte, ¿cómo habría afectado tu desempeño? & Impact of AI on second part & Open-ended \\
\hline
Q17 & Describe tu percepción general sobre el examen: ¿Cómo fue tu experiencia? & General perceptions of the exam format & Open-ended \\
\hline
Q18 & ¿Al comparar tu desempeño durante las dos sesiones, con y sin IA, que conclusión obtienes? & Comparison with and without AI & Open-ended \\
\hline
Q19 & ¿Crees que el examen evaluó tus conocimientos y habilidades en programación? ¿Por qué? & Exam really assessed programming skills & Open-ended \\
\hline
Q20 & ¿Qué sugerencias tienes para mejorar el examen? & Suggestions to improve the exam & Open-ended \\
\hline
Q21 & ¿Crees que la IA es una herramienta útil para aprender a programar? & AI useful for learning programming & Likert scale \\
\hline
Q22 & Con la llegada de herramientas de IA que generan código automáticamente, ¿cuál de las siguientes opciones describe mejor tu opinión sobre aprender a programar? & AI replaces coding skills & Multiple choice \\
\hline
Q23 & ¿Cómo te gustaría utilizar la IA en el futuro para aprender o programar? & Future use of AI in programming & Open-ended \\
\hline
Q24 & ¿Cómo te gustaría que se incorporará la IA en el futuro, en los cursos de programación en la Universidad? & AI role in programming courses & Open-ended \\
\hline

\end{tabularx}
}
\end{table}

\clearpage
\section{Sample Qualitative Responses}
\label{appendix:responses}

\begin{table}[h!]
\centering
\scriptsize
\renewcommand{\arraystretch}{1.4}
\setlength{\tabcolsep}{6pt}
\caption{Representative responses to Q8 in original Spanish and English translation}
\begin{tabular}{|p{2cm}|p{4cm}|p{5cm}|}
\hline
\textbf{Category} & \textbf{Original Quote (Spanish)} & \textbf{English Translation} \\
\hline
Clear Understanding & “Sí, porque comprendía toda la programación básica que se aplica, la sintaxis de Java es lo que me confunde un poco pero no mucho porque ya conozco C++ y es muy similar.” & “Yes, because I understood all the basic programming being applied; Java syntax confused me a bit, but not much since I already know C++, which is very similar.” \\
\hline
Partial or Conditional Understanding & “Entendía una gran parte del código, pero había algunas cosas de la sintaxis que me costaba entender.” & “I understood a large part of the code, but there were some parts of the syntax that were difficult to grasp.” \\
\hline
Use of AI to Request Clarifications & “Sí, ya que le pedía explicación del código generado.” & “Yes, since I asked it to explain the code it generated.” \\
\hline
AI Output Required Adaptation & “Repasándolo un poco se podía entender casi todo… pero no era exactamente como lo veíamos en clase.” & “By reviewing it a bit, almost everything could be understood… but it wasn’t exactly how we saw it in class.” \\
\hline
Minimal Comprehension Focus & “Realmente no tanto, debido a que siendo el objetivo crear el programa y que este funcionase, no me detuve a entender parte por parte.” & “Not really, because the goal was to get the program working, I didn’t stop to understand it piece by piece.” \\
\hline
\end{tabular}
\label{tab:q8-appendix}
\end{table}

\begin{table}[h!]
\centering
\scriptsize
\renewcommand{\arraystretch}{1.4}
\setlength{\tabcolsep}{6pt}
\caption{Representative responses to Q9 in original Spanish and English translation}
\begin{tabular}{|p{2cm}|p{4cm}|p{5cm}|}
\hline
\textbf{Theme} & \textbf{Original Quote (ES)} & \textbf{English Translation} \\
\hline
Refining and restructuring & Cambie algunos nombres de variables y algunos textos que aparecerían posteriormente en pantalla. & I changed variable names and texts that would later appear on screen. \\
\hline
Simplifying AI solutions & Me di cuenta que se podía hacer más corto eso y lo modifiqué. & I realized I could make it shorter and I modified it. \\
\hline
Extending or improving AI output & Introduje las imágenes de mi preferencia, y opciones nuevas. & I added images of my choice and new options. \\
\hline
Error-driven adaptation & No generé el código a partir de IA, sino que la use para corregir errores y aprender más acerca de la Programación Orientada a Objetos. & I didn’t generate code with AI, I used it to correct errors and learn more about object-oriented programming. \\
\hline
Minimal or no adaptation & No, ya que solo le pedi que me indicara errores en el código y que me diera librerías. & No, I only asked it to show me the errors and give me the libraries. \\
\hline
\end{tabular}
\end{table}

\begin{table}[htbp]
\centering
\scriptsize
\renewcommand{\arraystretch}{1.4}
\setlength{\tabcolsep}{6pt}
\caption{Representative responses to Q10 in original Spanish and English translation}
\label{tab:q10_app}
\begin{tabular}{|p{2cm}|p{4cm}|p{5cm}|}
\hline
\textbf{Category} & \textbf{Original Quote (Spanish)} & \textbf{English Translation} \\
\hline
High confidence & Me sentí confiado ya que soluciona problemas muy fácilmente y tenía claridad sobre mi código & I felt confident because it solves problems very easily and I had clarity about my code. \\
\hline
Moderate confidence with doubts & Al principio me sentí bastante confiado al usar el código generado por la IA, [...] me quedó la duda de si había algún error oculto & At first I felt confident using the AI-generated code, [...] but I had some doubts—whether it was efficient or had hidden errors. \\
\hline
Partial confidence or skepticism & Más o menos, en ocasiones usaba el código recién generado de la IA [...] usaba conceptos que nunca había visto yo antes & Sometimes I used the code and it worked perfectly, but other times it generated errors or used unfamiliar concepts I didn’t understand. \\
\hline
Low confidence or distrust & No me sentí confiado, siempre me detuve a revisar lo que la IA generaba [...] me daba soluciones erróneas & I didn’t feel confident, I always stopped to check the code because the AI often gave me wrong solutions. \\
\hline
\end{tabular}
\end{table}

\begin{table}[htbp]
\centering
\scriptsize
\renewcommand{\arraystretch}{1.4}
\setlength{\tabcolsep}{6pt}
\caption{Representative responses to Q13 in original Spanish and English translation.}
\label{tab:q13-appendix}
\begin{tabular}{|p{2cm}|p{4cm}|p{5cm}|}
\hline
\textbf{Category} & \textbf{Original (Spanish)} & \textbf{English Translation} \\
\hline
Conceptual scaffolding & Usar la IA me dio un soporte para luego poder realizar modificaciones al código por mi mismo. & Using AI gave me a foundation so I could later make modifications to the code by myself. \\
\hline
Code reusability & Había cosas dentro del código que podía reciclar para resolver los nuevos problemas. & There were things in the code I could reuse to solve the new problems. \\
\hline
Debugging preparedness & Me ayudó un poco para prepararme frente a posibles errores en el programa más adelante. & It helped me a bit to prepare for possible errors in the program later on. \\
\hline
Partial understanding & Era un poco más sencillo declarar las funciones y constructores como lo había planteado la IA, pero igualmente… fue un poco complicado. & It was a bit easier to declare functions and constructors as the AI had suggested, but even so... it was a bit complicated. \\
\hline
No benefit & No, creo que la IA al contrario, me facilitó demasiado la primera parte, lo que hizo que la segunda fuera un verdadero reto. & No, I think the AI actually made the first part too easy, which made the second part a real challenge. \\
\hline
\end{tabular}
\end{table}

\begin{table}[htbp]
\centering
\scriptsize
\renewcommand{\arraystretch}{1.4}
\setlength{\tabcolsep}{6pt}
\caption{Representative responses to Q14 in original Spanish and English translation}
\label{tab:q14_app}
\begin{tabular}{|p{2cm}|p{4cm}|p{5cm}|}
\hline
\textbf{Category} & \textbf{Original Quote (Spanish)} & \textbf{English Translation} \\
\hline
High confidence & Si, porque comprendí los conceptos requeridos y tenía una imagen mental de cómo hacerlo & Yes, because I understood the concepts and had a mental image of how to do it. \\
\hline
Moderate/conditional confidence & Más o menos, me sentí cómodo porque creí "saber" cómo funcionaba mi programa. & More or less, I felt comfortable because I thought I knew how my program worked. \\
\hline
Low confidence & NO, no tengo las herramientas del lenguaje para modificarlo no sé java & No, I don’t have the tools to modify it—I don’t know Java. \\
\hline
Mixed or evolving confidence & Un poco, debido a que ya sabía el que hacía el código, pero algunas cosas no sabía cómo funcionaban & A bit, because I knew what the code did, but there were things I didn’t know how they worked. \\
\hline
\end{tabular}
\end{table}

\begin{table}[htbp]
\centering
\scriptsize
\renewcommand{\arraystretch}{1.4}
\setlength{\tabcolsep}{6pt}
\caption{Representative responses to Q15 in original Spanish and English translation}
\label{tab:q15_app}
\begin{tabular}{|p{2cm}|p{4cm}|p{5cm}|}
\hline
\textbf{Category} & \textbf{Original Quote (Spanish)} & \textbf{English Translation} \\
\hline
Syntax and language-specific issues & Creo que entender las funciones de Java, entender la correcta sintaxis fue mi mayor dificultad & I think understanding Java functions and the correct syntax was my biggest difficulty. \\
\hline
Debugging and error-handling & La mayor fue la falta de una referencia inmediata para verificar si estaba en el camino correcto [...] & The biggest issue was not having a reference to check if I was on the right track, especially when I had doubts about the code. \\
\hline
Conceptual gaps & Diría yo que el entendimiento de los conceptos más “básicos” [...] uno tiene la ilusión de que entendió & You think you understood how the code worked, but trying to create it on your own shows you didn’t fully get it. \\
\hline
Task-specific challenges & De pronto me pareció un poco difícil lo de Restaurar Cartelera, en la parte de hacer el Backup & I found the part about restoring the billboard, especially the backup part, a bit difficult. \\
\hline
No major difficulty & Ninguna, simplemente el tiempo que me tomara realizarlo & None, just the time it took to do it. \\
\hline
\end{tabular}
\end{table}

\begin{table}[htbp]
\centering
\scriptsize
\renewcommand{\arraystretch}{1.4}
\setlength{\tabcolsep}{6pt}
\caption{Representative responses to Q16 in original Spanish and English translation}
\label{tab:q16_app}
\begin{tabular}{|p{2cm}|p{4cm}|p{5cm}|}
\hline
\textbf{Category} & \textbf{Original Quote (Spanish)} & \textbf{English Translation} \\
\hline
Increased efficiency and speed & Habría sido muchísimo más fácil porque optimizamos el tiempo que se demora uno revisando [...] & It would’ve been much easier because we would’ve saved time figuring out where and how to do something. \\
\hline
Confidence and error reduction & Me habría ayudado a depurar errores y optimizar el código, lo que habría mejorado mi desempeño & It would have helped me debug errors and optimize the code, improving my performance. \\
\hline
No significant impact & En nada, hubiera sido exactamente el mismo & It would’ve been exactly the same. \\
\hline
Trade-off: speed vs. learning & Siento que atrasa mucho el permitirnos desarrollar las habilidades de recursividad y lógica [...] & AI would have helped with time, but delays the development of problem-solving and logic skills. \\
\hline
\end{tabular}
\end{table}

\begin{table}[htbp]
\centering
\scriptsize
\renewcommand{\arraystretch}{1.4}
\setlength{\tabcolsep}{6pt}
\caption{Representative responses to Q17 in original Spanish and English translation}
\label{tab:q17_app}
\begin{tabular}{|p{2cm}|p{4cm}|p{5cm}|}
\hline
\textbf{Category} & \textbf{Original Quote (Spanish)} & \textbf{English Translation} \\
\hline
Positive and well-structured & Me gustó el formato de parcial [...] mide nuestras capacidades para modificar un código [...] & I liked the format... it tests our ability to modify a code that could have been made 100\% by AI. \\
\hline
Reinforced learning & Mi experiencia con el examen fue bastante desafiante pero útil [...] & It was a challenging but useful experience... it was a good opportunity to apply what I had learned. \\
\hline
Stressful or demanding & [...] la segunda parte fue un poco más estresante ya que se debía pensar en cómo hacer la modificación [...] & [...] the second part was a bit more stressful since we had to modify the code just by looking at it. \\
\hline
Neutral or uneventful & Buena, tuve una buena sensación con el examen. & Good, I had a good feeling about the exam. \\
\hline
Suggestions for improvement & Tediosa, es preferible acabar el parcial en una sola clase, como normalmente es. & It’s preferable to finish the exam in a single session, as usual. \\
\hline
\end{tabular}
\end{table}

\begin{table}[htbp]
\centering
\scriptsize
\renewcommand{\arraystretch}{1.4}
\setlength{\tabcolsep}{6pt}
\caption{Representative responses to Q18 in original Spanish and English translation}
\label{tab:q18_app}
\begin{tabular}{|p{2cm}|p{4cm}|p{5cm}|}
\hline
\textbf{Category} & \textbf{Original Quote (Spanish)} & \textbf{English Translation} \\
\hline
Increased efficiency and confidence with AI & Al comparar mi desempeño, noté que con la IA pude avanzar más rápido y con más confianza... & With AI I was able to move faster and feel more confident... \\
\hline
AI hinders deep understanding & Es radicalmente distinta la experiencia [...] termina siendo perjudicial [...] uno cree que ya las sabe, cuando no es así. & The experience is radically different [...] it ends up being harmful [...] you think you know it, but you don’t. \\
\hline
Revealed areas for improvement & En mi caso, debo aprender a organizar mejor el código y practicar más para poder resolver rápido [...] & I need to better organize my code and practice more to solve problems quickly... \\
\hline
AI as preparation for autonomy & Que en la parte con IA resolví dudas que me dejaron preparado para la segunda. & In the part with AI I resolved doubts that prepared me for the second. \\
\hline
Neutral or no clear conclusion & ninguna, no tengo ninguna conclusión. & none, I don’t have any conclusion. \\
\hline
\end{tabular}
\end{table}

\begin{table}[htbp]
\centering
\scriptsize
\renewcommand{\arraystretch}{1.4}
\setlength{\tabcolsep}{6pt}
\caption{Representative responses to Q19 in original Spanish and English translation}
\label{tab:q19_app}
\begin{tabular}{|p{2cm}|p{4cm}|p{5cm}|}
\hline
\textbf{Category} & \textbf{Original Quote (Spanish)} & \textbf{English Translation} \\
\hline
Yes, comprehensive assessment & Sí, creo que el examen evaluó bien mis conocimientos y habilidades en programación [...] & Yes, I believe the exam assessed my programming knowledge and skills well [...] \\
\hline
Partially assessed & Más o menos, porque si evaluó mis conocimientos, pero siento que no tanto mis habilidades. & More or less, because it evaluated my knowledge but not so much my skills. \\
\hline
No, limited assessment & NO, solo en la redacción de promps y conocimiento general de la logica de programacion. & No, only in prompt writing and general programming logic. \\
\hline
\end{tabular}
\end{table}

\begin{table}[htbp]
\centering
\scriptsize
\renewcommand{\arraystretch}{1.4}
\setlength{\tabcolsep}{6pt}
\caption{Representative responses to Q20 in original Spanish and English translation}
\label{tab:q20_app}
\begin{tabular}{|p{2cm}|p{4cm}|p{5cm}|}
\hline
\textbf{Category} & \textbf{Original Quote (Spanish)} & \textbf{English Translation} \\
\hline
Adjust timing & Que se haga en una sola clase. & That it should be done in a single session. \\
\hline
Better guidance & Tener mayor claridad en los requerimientos de cada ejercicio. & Provide greater clarity on the requirements of each exercise. \\
\hline
Improve exam design & Que la primera parte sea con IA pero de manera restringida. & That the first part should include AI, but in a more restricted way. \\
\hline
More preparation or practice & Más familiarización previa del Lenguaje con el que se va a trabajar. & More prior familiarization with the language that will be used. \\
\hline
No suggestions & Ninguna. & None.\\
\hline
Fairness or grading issues & No generar tantas preguntas extras [...] los estudiantes a los cuales se les calificó primero tienen pocos métodos. & Avoid too many extra questions [...] students who were graded first had fewer methods.\\

\hline
\end{tabular}
\end{table}

\begin{table}[htbp]
\centering
\scriptsize
\renewcommand{\arraystretch}{1.4}
\setlength{\tabcolsep}{6pt}
\caption{Representative responses to Q23 in original Spanish and English translation}
\label{tab:q23_app}
\begin{tabular}{|p{2cm}|p{4cm}|p{5cm}|}
\hline
\textbf{Category} & \textbf{Original Quote (Spanish)} & \textbf{English Translation} \\
\hline
Support for learning and debugging & Me gustaría usar la IA como una herramienta de apoyo para resolver dudas, depurar código y obtener sugerencias. & I would like to use AI as a support tool to resolve doubts, debug code, and get suggestions. \\
\hline
Selective or cautious use & Para resolver ciertos problemas que se me compliquen, pero no para que me haga todas las cosas. & To solve problems that are difficult for me, but not to do everything for me. \\
\hline
Automation and optimization & Para optimizar el tiempo en proyectos grandes, que la IA se encargue de la parte más sencilla. & To optimize time in large projects, letting AI handle the simplest parts. \\
\hline
Minimal or no use & Preferiría aprenderlo totalmente por mi cuenta para que al final solo sea una mera herramienta más. & I would rather learn entirely on my own so that in the end it’s just another tool. \\
\hline
Learning from AI explanations & Pidiéndole que explique procesos, o resuma documentaciones. & Asking it to explain processes or summarize documentation. \\
\hline
\end{tabular}
\end{table}

\begin{table}[htbp]
\centering
\scriptsize
\renewcommand{\arraystretch}{1.4}
\setlength{\tabcolsep}{6pt}
\caption{Representative responses to Q24 in original Spanish and English translation}
\label{tab:q24_app}
\begin{tabular}{|p{2cm}|p{4cm}|p{5cm}|}
\hline
\textbf{Category} & \textbf{Original Quote (Spanish)} & \textbf{English Translation} \\
\hline
Support tool & Me gustaría que la IA se usara en los cursos de programación para ayudarnos a resolver dudas, depurar código y dar explicaciones más claras. & I would like AI to be used in programming courses to help us solve doubts, debug code, and give clearer explanations. \\
\hline
Autonomous use & Que sea parte del aprendizaje autónomo, más no una herramienta para la enseñanza. & It should be part of autonomous learning, but not a tool for teaching. \\
\hline
No AI in courses & De ninguna manera, manteniendo una buena pedagogía y abarcando correctamente los conocimientos. & Not at all, maintaining good pedagogy and fully covering the concepts. \\
\hline
Understanding AI systems & Me gustaría saber cómo funcionan y crear una. & I would like to learn how they work and create one. \\
\hline
\end{tabular}
\end{table}

\end{document}